\documentclass[a4paper,11pt]{article}
\pdfoutput=1
\usepackage{jheppub}
\usepackage[normalem]{ulem}
\usepackage{xcolor}

\newcommand{\ii}{\text{i}}

\newcommand{\dd}[1]{\text{d}#1 \, }
\newcommand{\e}[1]{\text{e}^{#1}}
\newcommand{\im}{\text{Im} \,}
\newcommand{\rH}{r_{\text{H}}}

\newcommand{\violet}{\textcolor{violet}}
\newcommand{\cyan}{\textcolor{cyan}}
\newcommand{\blue}{\textcolor{blue}}
\newcommand{\green}{\textcolor{green}}
\newcommand{\olive}{\textcolor{olive}}
\newcommand{\orange}{\textcolor{orange}}
\newcommand{\red}{\textcolor{red}}

\title{Horizon Tunneling Revisited:\\
The Case of Higher Dimensional Black Holes}

\author{Madhusudhan Raman}
\affiliation{Institute of Mathematical Sciences\\
Homi Bhabha National Institute (HBNI)\\
IV Cross Road, C.~I.~T.~Campus\\
Taramani, Chennai 600 113\\
Tamil Nadu, India}
\emailAdd{madhur@imsc.res.in}

\abstract{We study the tunneling of massless scalars across black hole horizons in any number of spacetime dimensions greater than three. Our analysis finds that corrections due to backreaction and the inverse dimensional expansion are naturally concomitant, and furnishes a simple proof of the classic relation between entropy and area in all spacetime dimensions, finite or infinite. We conclude with a discussion of the limit in which the the number of spacetime dimensions is taken to infinity, where we find that thermodynamic quantities are related to the ``thickness'' of the membrane on which all the curvature is localized.}

\begin{document} 
\maketitle
\flushbottom

\section{Motivations}
\label{sec:Intro}

\subsection*{The Large-D Paradigm}
A recent and intriguing development in understanding gravity has been the resurrection of the large-$D$ paradigm in general relativity, introduced by \cite{Strominger:1981jg}. While there has been sustained interest in studying gravity in greater than four dimensions (see \cite{Emparan:2008eg} for a comprehensive review of this line of inquiry), an interest in the limit $D \rightarrow \infty$ was sparked by \cite{Emparan:2013moa}, where the authors argued that in this limit, a natural separation of length scales native to the problem allowed for a surprisingly efficient study of scalar wave absorption and the black brane instability.

Since then, a large body of work discussing various aspects of black holes in large-$D$ has appeared, focusing on the dramatic simplifications that arise as in this limit \cite{Emparan:2013xia,Emparan:2014aba,Emparan:2014cia,Emparan:2014jca,Bhattacharyya:2015dva,Emparan:2015hwa,Suzuki:2015iha,Emparan:2015rva,Tanabe:2016opw,Bhattacharyya:2015fdk,Dandekar:2016fvw,Bhattacharyya:2016nhn}. Of particular note, \cite{Bhattacharyya:2015dva,Emparan:2015hwa} made use of the poignant observation that in the large-$D$ limit, any curvature is \emph{strongly localized near the horizon}, and that spacetime quickly becomes flat outside of it: in effect, a ``membrane''. This observation, coupled with the aforementioned separation of scales allowed for the development of effective theories describing the dynamics of this membrane. The membrane picture has, since then, been generalized to charged membranes \cite{Bhattacharyya:2015fdk}, and extended to subleading order in the inverse-dimensional expansion \cite{Dandekar:2016fvw,Bhattacharyya:2016nhn}. These developments have been called a \emph{membrane paradigm}.

These analyses hope to better understand the dynamics of black holes in a handful (or two!) of spacetime dimensions by studying the limit in which the number of spacetime dimensions is infinite. Ostensibly, one would go about this by computing all quantities as expansions in a series in $1/D$, then plugging in the dimension of interest, and reading off the answer. In this work, we will address the question of whether thermodynamic quantities like entropy can be straightforwardly computed in this manner. As a check of our computations, we will also ask whether the resulting series can be resummed to recover the celebrated relation between the area of a black hole and its entropy. For definiteness, we will restrict ourselves to a study of the so-called Schwarzschild-Tangherlini black hole, whose metric in Schwarzschild coordinates is:
\begin{equation}
\label{eq:TangBH}
\dd{s^2} = -\left( 1- \left(\frac{r_\text{H}}{r}\right)^{d}\right) \,\dd{t^2} + \left( 1- \left(\frac{r_\text{H}}{r}\right)^{d}\right)^{-1} \, \dd{r^2} + r^2 \, \dd{\Omega_{d+1}} \ .
\end{equation}
Here, $r_{\text{H}}$ is the horizon radius, and $d \geq 1$. The number $d$ is related to the dimension of spacetime as
\begin{equation}
D = d+3 \ .
\end{equation}

Our interest in this program was prompted by the effective membrane-like quality of infinite dimensional black holes, an idea developed in \cite{Emparan:2013moa} and described in the following manner in \citep{Bhattacharyya:2015dva}. Consider hovering just outside the horizon $r_{\text{H}}$ at some fixed distance $r$ and subsequently consider the limit $d \rightarrow \infty$. In this case,
\begin{equation}
\left(\frac{r_\text{H}}{r}\right)^{d} \rightarrow 0 \ ,
\end{equation}
and the metric \eqref{eq:TangBH} reduces to the metric of flat spacetime. This way of taking the limit is not very enlightening, but there is a more interesting alternative. This time around, let us consider hovering just outside the horizon at a distance $r$, except this time, let 
\begin{equation}
r = r_{\text{H}} \left( 1 + \frac{R}{d} \right) \ .
\end{equation}
In the above expression, $R$ multiplies a quantity we will refer to as the \emph{thickness}
\begin{equation}
\ell = \frac{r_{\text{H}}}{d} \ ,
\end{equation}
which in turn means
\begin{equation}
r - r_{\text{H}} \simeq \ell \ .
\end{equation}
The thickness here is a measure of the extent to which curvature is localized near $\rH$. It is easy to see that in the limit $D \rightarrow \infty$, the thickness goes to zero inversely as the dimension of spacetime. Now, we hold $R$ fixed as we take $d \rightarrow \infty$. In this case, 
\begin{equation}
\lim_{d\rightarrow \infty} \left(\frac{r_\text{H}}{r}\right)^{d} = \e{-R} \ .
\end{equation}
The lesson here is that in the limit $d \rightarrow \infty$, the region surrounding the black hole that feels its presence i.e.~the region that can interact with it gravitationally is O$(\ell)$ from its horizon. Outside this membrane the spacetime is flat. The take-away from this discussion is that a natural length-scale associated to an infinite-dimensional black hole is $\ell$.

\subsection*{Hawking Processes}
Turning to our other source of motivation, an intuitive explanation for why black holes radiate (\cite{Hawking:1974sw}, see \cite{Jacobson:2003vx} for a nice review) is the stretching of spacetime near horizons, discussed in detail in \cite{Mathur:2008wi}. We briefly review this argument below.\footnote{This argument was described to us by Bidisha Chakrabarty at ST$^4$ 2017, and we thank her for a patient and lucid discussion of the same.}

Let us imagine starting with a free quantum field in a curved spacetime, whose Fourier modes behave like harmonic oscillators with some frequency $\omega$. Over time, it is possible that due to changes in the underlying spacetime, the frequency of this Fourier mode will change to $\omega'$. If this change occurs adiabatically and we begin in a vacuum state, the adiabatic theorem guarantees that we will remain in a vacuum state.\footnote{By vacuum state we mean a particular Fourier mode being unoccupied.} If, however, this change is sudden, a vacuum state for a harmonic well with frequency $\omega$ will find that it is no longer a vacuum state for a harmonic well with frequency $\omega'$, but instead an excited state that should be understood as a superposition of eigenstates of the new harmonic well. That is, an erstwhile vacuum will find itself populated with particles.

Slow and fast changes here are distinguished by appealing to the only time scale in the problem: the frequencies of the wells. That is, when $\omega$ and $\omega'$ differ by a small amount, the only time scale in the problem is
\begin{equation}
\delta \sim \frac{1}{\omega} \ ,
\end{equation}
and any determination of (a)diabaticity should be made with reference to the time scale $\delta$.

Now, massless particles in a black hole spacetime will travel along null geodesics. For the moment, let us only consider radial motion. In the $t$-$r$ plane, null geodesics ``peel away'' from the horizon, i.e.~null geodesics on both sides of the horizon move away from $\rH$ at later times. This \emph{stretching} of spacetime in the neighbourhood of the horizon leads to particle production in its vicinity: the Hawking process. In turn, the radiation spectrum of the black hole can be described by a temperature (that is related to its surface gravity). Other thermodynamic quantities also find natural analogues \cite{Wald:1999vt,Carlip:2014pma}.

We now combine these arguments as follows: such a stretching of the spacetime near the horizon would, as the dimension of the black hole spacetime is increased, be more and more strongly localized near $r_{\text{H}}$. We would like to understand what consequences this localization of curvature at the horizon has for thermodynamic quantities---specifically, the entropy and the temperature---by computing them in an inverse-dimensional expansion, thereby allowing for the discussion of both finite- and infinite-dimensional black holes within the same framework. We will find that these thermodynamic quantities are determined by the thickness of the membrane, or the extent to which most of the curvature is localized near the horizon. We now turn to the third theme of our study, which will also afford us an opportunity to discuss the methods we will employ.

\subsection*{Backreaction}
One of the features of the semi-classical analysis \cite{Hawking:1974sw} is that it applies only to stationary black holes. Intuitively, this is problematic because a black hole radiating would cause it to lose mass, thereby rendering the spacetime non-stationary; further, a failure to account for this results in violations of energy conservation. A more thorough analysis of Hawking radiation should take into account the fact that the black hole metric changes as Hawking quanta are radiated: this is called the backreaction problem. Although it has not been satisfactorily solved, steps towards take into account backreaction were taken in \cite{Kraus:1994by,Kraus:1994fj}. An outgrowth of these studies yielded the \emph{tunneling} or \emph{null-geodesic} method, developed by \cite{Srinivasan:1998ty,Parikh:1999mf,Shankaranarayanan:2000gb,Padmanabhan:2004tz}, and reviewed in \cite{Vanzo:2011wq}. We will follow the work of Parikh-Wilczek \cite{Parikh:1999mf} in our own study. 

The null-geodesic method relates the probability $\Gamma$ of a particle tunneling across the horizon to the change in entropy of a black hole:
\begin{equation}
\Gamma \sim e^{\Delta S_{\text{BH}}} \ ,
\end{equation}
thus giving traction to the popular heuristic that Hawking quanta ``come from'' tunneling across the horizon. This tunneling probability is computed within the WKB approximation, where we expect
\begin{equation}
\Gamma \sim e^{-2 \, \im S} \ ,
\end{equation}
Here, $S$ is the action of the particle. It is important to keep in mind that when the radiated Hawking quantum is traced back to the horizon, it is infinitely blueshifted. That is, a wave with finite frequency far away from the black hole will appear extremely energetic near the horizon. Thus, we are justified in treating it within the WKB framework. 

Backreaction in this context is taken into account in the following way: let us say a particle with energy $\omega$ has tunneled out of the black hole. Its motion is governed by null geodesics in a now \emph{modified} gravitational background, where the mass $M$ of the black hole is reduced to $(M - \omega)$. Of particular note, accounting for leading-order backreaction in this way has a happy consequence: the emission spectrum of a black hole can be shown \emph{not} to be purely thermal.

In higher-dimensional spacetimes, the gravitational potentials fall off faster, rendering gravity weaker. We expect as a result that as we dial up the dimension of spacetime, the effects of backreaction become less important. In the strict large-$D$ limit, backreaction should therefore be negligible. Alternatively, we may say that backreaction is suppressed in the $1/D$ expansion. Incorporating leading-order backreaction and organizing the expressions in a $1/D$ expansion also gives us a series in $\omega$, the energy of the radiated particle. In addition to generalizing the Parikh-Wilczek analysis to arbitrary dimensions, our main result in this paper is that the series that accounts for leading-order backreaction is the same as the inverse-dimensional expansion.

\subsection*{Organization}
Our main object of interest will be the imaginary part of the classical action $S_0$. We will organize it as a series in inverse powers of the spacetime dimension, while simultaneously incorporating leading-order backreaction, which also takes the form of a power series. All of this is at leading-order in $\hbar$. From this, we intend to read off the temperature and entropy of the black hole. We will review the details of how we do this as we go along.

Section \ref{sec:BH} generalizes the Gullstrand-Painlev\'e coordinate system to $D$ spacetime dimensions, thus giving us a coordinate system that is smooth in the vicinity of the horizon and allowing us to meaningfully discuss tunneling across horizons. We also determine radial null geodesics in these spacetimes, as these will be the trajectories followed by massless particles. Section \ref{sec:Tunnel} focuses on computing the imaginary part of the action, and also considers subleading (in $1/D$) corrections to the Bekenstein-Hawking entropy in the limit $D \rightarrow \infty$. This section finds that backreaction can be thought of as a series in the energy of the radiated particle, thus giving us two series to contend with: backreaction and the inverse-dimensional expansion. We find that they are naturally paired. In Section \ref{sec:kSubLead} we find that the integrals in question can be done exactly---effectively, a resummation of the series in $1/D$---and for all black holes (finite- or infinite-dimensional) we are able to show that our results reproduce the classic relation between entropy and area. In Section \ref{sec:LargeD}, we consider the limit in which the number of spacetime dimensions is sent to infinity. We find that our analysis carries over rather naturally in this limit; in each case---whether the spacetime dimension is finite or infinite---the entropy of the black hole is expressed in terms of a natural length scale associated with the black hole. We then conclude with some future directions for research. Appendix \ref{sec:AppA} demonstrates that if one insisted on working in order-by-order in the large-$D$ limit, the answers one gets match the exact evaluation of the integral determining the imaginary part of the classical action. This is of course to be expected, but doing things this way teaches us a valuable lesson about expanding in the ``right'' parameter, in addition to justifying the applicability of the results quoted in Section \ref{sec:kSubLead} even in the limit $D \rightarrow \infty$.

\acknowledgments{We are happy to thank Sujay Ashok, Dileep Jatkar, Renjan John, Alok Laddha, Arunabha Saha, Ronak Soni, and Nemani Suryanarayana for encouragement, many discussions, and helpful comments on an earlier version of this preprint.}

\section{Black Hole Horizons}
\label{sec:BH}
In this section, we introduce the Schwarzschild-Tangherlini black hole \eqref{eq:TangBH}, change coordinates to a version of the Gullstrand-Painlev\'e coordinate system which is smooth across the horizon, and compute its radial null geodesics.

\subsection{Coordinates}
\label{sec:Coord}
The metric of a stationary, spherically symmetric black hole in $D = d+3$ spacetime dimensions in the usual Schwarzschild coordinates \cite{Tangherlini:1963bw} reads
\begin{equation}
\label{eq:SchwarzTang}
\dd{s^2} = -f \,\dd{t^2} + f^{-1} \, \dd{r^2} + r^2 \, \dd{\Omega_{d+1}} \ .
\end{equation}
with
\begin{equation}
f = 1- \left(\frac{r_\text{H}}{r}\right)^{d} \ .
\end{equation}
In the above expression, $r = r_\text{H}$ is the horizon radius. It is related to the mass $M$ as
\begin{equation}
\label{eq:rHM}
r_{\text{H}} [M] = \left( \frac{16\pi\, M}{(d+1)\,\Omega_{d+1}} \right)^{1/d} \ .
\end{equation}
Here $\Omega_{d+1}$ is the area of a unit $(d+1)$-sphere:
\begin{equation}
\label{eq:2SphereArea}
\Omega_{d+1} = \frac{2\pi^{\frac{d+2}{2}}}{\Gamma(\frac{d+2}{2})} \ ,
\end{equation}
It is easy to check that $r_{\text{H}} = 2M$ in $d = 1$ (equivalently, $D = 4$). Our strategy, consistent with the membrane paradigm, will be to first fix the horizon radius, even for finite-dimensional black holes. This is in contrast with the usual practice in general relativity, where the mass of the black hole is taken to uniquely describe a Schwarzschild spacetime. We adopt the convention that $\rH$ (when written without an argument) is this fixed horizon radius, i.e.
\begin{equation}
\rH := \rH[M] \ ,
\end{equation}
and we choose to suppress the dependence of $r_{\text{H}}$ on $M$ in the interest of keeping the resulting expressions as short and clear as possible. This point will be important when we discuss the tunneling of particles across the black hole horizon; we expect that when a particle carries away some energy $\omega$, the mass of the black hole decreases to $(M-\omega)$, and consequently, the radius shrinks.\footnote{In the large-$D$ limit, as $M$ goes to zero semifactorially, so too will $\omega$ in such a way that the condition $\omega \leq M$ is always preserved. This is a consequence of energy conservation, and is in principle independent of dimensional considerations.} That is,
\begin{equation}
\label{eq:rHMOmega}
\rH' := \rH[M-\omega] \ ,
\end{equation}
where as before, $\rH[\ \cdot \ ]$ is defined in \eqref{eq:rHM}. Finally, the null-geodesic method we will shortly introduce focuses on the $t$-$r$ plane, so in all that follows we will suppress any dependence of the metric on angular variables, which in our context will be of the form $r^2 \, \dd{\Omega_{d+1}}$. This is acceptable as long as we are restricting ourselves to the analysis of $s$-wave emission.

In order to meaningfully discuss tunneling phenomena across the horizon, we must work in a coordinate system that is smooth in its vicinity. This requires that we generalize the Gullstrand-Painlev\'e coordinate system (originally adapted to $d=1$) by shifting $t$ as
\begin{equation}
t \rightarrow \widetilde{t} = t + g \ ,
\end{equation}
where $g$ is some function of the radial coordinate that is as yet undetermined. We have
\begin{align}
\dd{s^2} = -f \, \dd{\widetilde{t}^2} + \left( -f \left( g' \right)^2 + f^{-1} \right) \dd{r^2} + 2 f g' \, \dd{\widetilde{t}} \, \dd{r} \ .
\end{align}
We're going to insist that spatial slices in this coordinate system are flat, so we set the coefficient of $\dd{r^2}$ to unity, which yields
\begin{equation}
g' = \sqrt{\frac{1-f}{f^2}} \ ,
\end{equation}
and in turn gives us a metric that looks like
\begin{equation}
\label{eq:DSchwarzGP}
\dd{s^2} = -f \,\dd{\widetilde{t}^2} + 2 \sqrt{1-f} \, \dd{\widetilde{t}} \, \dd{r} + \dd{r^2} \ .
\end{equation}

\subsection{Radial Null Geodesics}
\label{sec:RadNull}
Since we will soon consider massless particles moving in the above background, we must consider geodesics that are both radial and null, i.e.~that $\left(\frac{\dd{s}}{\dd{t}}\right)^2 = 0$, which give
\begin{equation}
\label{eq:RadialNull}
\dot{r} = \frac{\dd{r}}{\dd{\widetilde{t}}} = \pm 1 - \sqrt{1-f} \ ,
\end{equation}
where $+$ ($-$) correspond to outgoing (respectively, incoming) geodesics. It will turn out that the momentum of the particle tunneling through the horizon is inversely related to $\dot{r}$.

\section{Tunneling}
\label{sec:Tunnel}
We consider a massless scalar field $\phi$ that is minimally coupled to our background metric \eqref{eq:DSchwarzGP}. Its equation of motion is the Klein-Gordon equation in curved spacetime. In keeping with the form of the WKB ansatz, we will write the field as
\begin{equation}
\label{eq:AnsatzWKBI}
\phi(x) = A(x) \, \e{\ii S(x)/\hbar} \ ,
\end{equation}
where we have separated the field into its amplitude $A(x)$ and its phase $S(x)$.

Qualitatively, this ansatz is uniquely poised to probe both kinds of fundamental motions executed by quantum mechanical systems, whether finite- or infinite-dimensional. If the phase is real, it will effect oscillatory (i.e.~\emph{perturbative}) motion; if the phase has an imaginary part, it will describe evanescent (i.e.~\emph{tunneling} or \emph{non-perturbative}) motion.

The WKB approximation requires that we assume the amplitude varies negligibly in comparison to the phase, so we update our ansatz to
\begin{equation}
\label{eq:AnsatzWKBII}
\phi(x) = A \, \e{\ii S(x)/\hbar} \ ,
\end{equation}
Finally, since we are working in a semi-classical picture, we expect that the usual results about tunneling amplitudes are still valid. In particular, we expect that the probability of a scalar particle tunneling across the horizon would be
\begin{equation}
\label{eq:TunnelingRate}
\Gamma \sim \text{exp}\left( -\frac{2}{\hbar} \, \im S\Big\vert^{x_f}_{x_i} \right) \ ,
\end{equation}
where $x_i$ and $x_f$ denote the beginning and end of the classically forbidden regions \cite{Parikh:2004ih}. As we have already argued, our use of the WKB approximation is justified by the infinite blueshift of the Hawking quantum close to the horizon.

The Klein-Gordon equation for a massless scalar field is
\begin{equation}
\hbar^2 \, g^{\mu \nu} \nabla_\mu \partial_\nu \, \phi = 0 \ .
\end{equation}
We now plug the ansatz \eqref{eq:AnsatzWKBII} into the above differential equation, and this gives us (on suppressing the dependence of the phase $S$ on the spacetime coordinates):
\begin{equation}
(\partial S)^2 - \ii \, \hbar \, \nabla^\mu \partial_\mu S = 0 \ .
\end{equation}
We now assume that the phase angle $S$ admits an expansion of the form
\begin{equation}
S = S_0 + \hbar \, S_1 + \hbar^2 S_2 + \cdots \ ,
\end{equation}
where we refer to $S_0$ as the classical action. We then organize the resulting differential equation order-by-order in $\hbar$. This yields an infinite family of differential equations, one for each order in $\hbar$. We will focus our attention on the leading contribution to the semi-classical limit, which yields the following equation of motion
\begin{align}
\label{eq:Hbar1}
\left( \partial S_0 \right)^2 &= 0 \ , 
\end{align}
whose solution we now turn to. In all that follows, we will take the liberty of setting $\hbar = 1$.

\subsection{Classical Action}
We will, following Parikh-Wilczek \cite{Parikh:1999mf}, work in the $s$-wave approximation, utilising fully the spherical symmetry of the gravitational background. In this approximation, the only relevant components of the inverse metric $g^{\mu\nu}$ are
\begin{align}
g^{\widetilde{t}\widetilde{t}} &= -1 \ , \\
g^{\widetilde{t} r} &= \sqrt{1-f} \ , \\
g^{rr} &= f \ .
\end{align}
We are now in a position to write down the equations of motion \eqref{eq:Hbar1}, which after some simple manipulations looks like
\begin{align}
\left( \partial_{\widetilde{t}} + \textcolor{blue}{\left( \pm 1 - \sqrt{1-f} \right)} \partial_r \right) S_0 &= 0 \ .
\end{align}
The expression highlighted above should be familiar: it controls radial, null geodesics \eqref{eq:RadialNull}. Thus, our differential equation for $S_0$ has simplified considerably, now reading
\begin{equation}
 \left( \partial_{\widetilde{t}} + \dot{r} \partial_r \right) S_0 = 0 \ ,
\end{equation}
which is solved by 
\begin{equation}
S_0 = \pm \omega \left(\widetilde{t} - \int^r \frac{\dd{r}}{\dot{r}} \right) \ ,
\end{equation}
as is straightforwardly checked. This object is the classical action for the massless scalar field in the semi-classical approximation, used implicitly by \cite{Parikh:1999mf}. Finally, since the time dependence is of a simple form $e^{\ii \omega \widetilde{t}}$, we choose to define a time-independent classical action as
\begin{equation}
\label{eq:S0Momentum}
S_0 = \int \dd{r} p(r) \ ,
\end{equation}
with 
\begin{equation}
p(r) = \frac{1}{\dot{r}} \ ,
\end{equation}
where $\dot{r}$ is given by \eqref{eq:RadialNull}.

\subsection{The Imaginary Part of the Action}
Consider a particle tunneling out of the horizon of our black hole. The imaginary part of its action is
\begin{align}
\im S_0 &= \im \int_{r_{\text{in}}}^{r_{\text{out}}} \dd{r} p(r) \ , \\
&= \im \int_{r_{\text{in}}}^{r_{\text{out}}} \dd{r} \int_{0}^{p(r)} \dd{p'} \ , \\
&= \im \int_{r_{\text{in}}}^{r_{\text{out}}} \dd{r} \int_{M}^{M-\omega} \frac{\dd{H}}{\dot{r}} \ ,
\end{align}
where in the last step we used Hamilton's equation to change the variable of integration from momentum to energy. The change in the limits of the integral reflects a crucial observation due to \cite{Kraus:1994by,Kraus:1994fj}, namely that if the total ADM mass is held fixed and the black hole mass is allowed to fluctuate---a precondition for Hawking processes---then a particle carrying an energy $\omega$ travels along geodesics given by the line element we derived earlier \eqref{eq:DSchwarzGP}, except that we must make the substitution 
\begin{equation}
r_{\text{H}} = r_{\text{H}} [M] \longrightarrow r_{\text{H}}' = r_{\text{H}} [M-\omega] \ .
\end{equation}
This substitution accounts for leading-order backreaction; the particle carries away some energy $\omega$ from the black hole, and this in turn causes the black hole mass to decrease by the same amount, by energy conservation. As before, the mass of our black hole will go to zero as we dial up the dimension of spacetime. However, this should not bother us, and we take $\omega$ to be some part of this vanishingly small mass as the principle of energy conservation is indifferent to this limit. As we argued in Section \ref{sec:Intro}, we expect that inverse powers of dimension and the corrections due to backreaction will appear in a concerted fashion. Thus, for the purposes of our analysis, we cannot afford to neglect it.

The energy $\omega$ of the Hawking quantum is some fraction of the ADM mass of the black hole. The $s$-wave approximation allows for the interpretation of this particle as akin to a spherically symmetric shell of mass $\omega$ thrown off by the black hole. We emphasise that this is an assumption, i.e.~we assume that a similar substitution is applicable in $D$ spacetime dimensions. On changing the variable of integration to $\omega' > 0$ corresponding to a positive energy particle tunneling out of the horizon, we get
\begin{align}
\im S_0 &= - \im \int_{r_{\text{in}}}^{r_{\text{out}}} \dd{r} \int_{0}^{\omega} \frac{\dd{\omega'}}{\dot{r}} \ , \\
&= - \im \int_{0}^{\omega} \dd{\omega'} \int_{r_{\text{in}}}^{r_{\text{out}}} \dd{r} \left[ 1- \left( \frac{r_{\text{H}}'}{r} \right)^{d/2} \right]^{-1} \ ,
\end{align}
where in the second line we have changed the order of integrals. Let us start with a specification of what the limits of integration over the radial coordinate are. We start with a particle right behind the \emph{old} horizon, at a small distance $\epsilon$. Further, as we discussed earlier, after tunneling the horizon itself shrinks to $r_{\text{H}}'$, so we'll want to stop integrating just outside the \emph{new} horizon. Hence, we make the choice
\begin{align}
r_{\text{in}} &= r_{\text{H}} - \epsilon \ , \\
r_{\text{out}} &= r_{\text{H}}' + \epsilon \ .
\end{align}
This gives us
\begin{equation}
\im S_0 = -\int_{0}^{\omega} \dd{\omega'} \textcolor{blue}{\left[\im \, \lim_{\epsilon \rightarrow 0} \int_{r_{\text{H}} - \epsilon}^{r_{\text{H}}' + \epsilon} \dd{r} \left[ 1- \left( \frac{r_{\text{H}}'}{r} \right)^{d/2} \right]^{-1} \right]} \ ,
\end{equation}
Let us tackle the highlighted expression that integrates over the radial coordinate first. The key is to only keep track of the imaginary part of this integral, whose sole contribution comes from the pole at $r = r_{\text{H}}'$. Armed with this intuition, we change variables
\begin{equation}
r = r_{\text{H}}' + s \ ,
\end{equation}
and expand about $s = 0$. This gives us
\begin{equation}
\left[ 1-\left( \frac{r_{\text{H}}'}{r_{\text{H}}'+ s} \right)^{d/2} \right]^{-1} \simeq + \frac{2 r_{\text{H}}'}{d} \frac{1}{s} + \cdots \ .
\end{equation}
The integral over $r$ now reduces to an integral over $s$, whose real part is its Cauchy principal value. Its imaginary part on the other hand evaluates to
\begin{equation}
-\frac{2 \pi \, r_{\text{H}}'}{d} \ .
\end{equation}
The factors of $(-1)$ compensate for each other, and we are left with an integral over $\omega'$; using \eqref{eq:rHM}, we must evaluate 
\begin{equation}
\label{eq:FiniteD}
\im S_0 = \frac{2\pi}{d} \int_{0}^{\omega} \dd{\omega'} r'_{\text{H}} \ ,
\end{equation}
where $\rH'$ is defined in \eqref{eq:rHMOmega}. This result was also derived by \cite{Wei:2014bva}.\footnote{We thank Pinaki Banerjee for bringing this work to our attention.} We will soon find that this integral can be evaluated exactly, and for all the higher-dimensional black holes we are considering. However, we pause here to check for consistency with known results.

\subsubsection{A Consistency Check}
When $d=1$, the imaginary part of the classical action becomes
\begin{equation}
\label{eq:ParikhWilczekResult}
\im S_0 = 4\pi \, \omega \left( M - \frac{\omega}{2} \right) \ ,
\end{equation}
Further,  via \eqref{eq:TunnelingRate} we compute the tunneling rate
\begin{align}
\label{eq:4DTunnel}
\Gamma \sim \exp\left( - 8 \pi M \, \omega + 4\pi\, \omega^2 \right) \ .
\end{align}
When we keep just the term linear in $\omega$, we may read off the inverse temperature of the black hole as $8\pi M$, consistent with \cite{Hawking:1974sw}. As emphasized by \cite{Parikh:1999mf}, energy conservation (by way of including the back-reaction) induces O$(\omega^2)$ corrections to the tunneling rate. To capture this, we may define an effective temperature
\begin{equation}
\frac{1}{T_{\text{eff}}} = 8 \pi \, M - 4\pi\, \omega \ ,
\end{equation}
and we see that the Hawking process effectively raises the temperature of the black hole as it radiates. Finally, we may appeal to the thermodynamic relation
\begin{equation}
\dd{\omega} = T \ \dd{S} \ ,
\end{equation}
whose integrated form allows us to read off the entropy of the black hole
\begin{align}
S &= 4 \pi \, M^2 \ , \\
&= \frac{A}{4} \ ,
\end{align}
which is consistent with semi-classical gravity calculations. Further, the factor in the exponential in \eqref{eq:4DTunnel} matches the change in the black hole entropy, as evaluated by \cite{KeskiVakkuri:1996xp,Massar:1999wg}. We will now tell the Parikh-Wilczek story in all higher dimensional black hole spacetimes.

\subsection{Tunneling in Higher Dimensions}
We have to evaluate \eqref{eq:FiniteD}, which we reproduce below, in $D$ spacetime dimensions. 
\begin{align}
\im S_0 &= \ \frac{2\pi}{d} \int_{0}^{\omega} \dd{\omega'} r'_{\text{H}} \ , \\
\label{eq:rHMomega}
&= \ \frac{2\pi}{d} \int_{0}^{\omega} \dd{\omega'} r_{\text{H}}[M-\omega']  \ ,
\end{align}
and---just for the moment---we will expand the reduced horizon radius $\rH'$ in the quantum of energy that the black hole radiates.\footnote{There are ways of evaluating this exactly, and without an expansion in inverse powers of the spacetime dimension; we do this because our interest is in studying the interplay between backreaction and the inverse-dimensional expansion.} We then get
\begin{align}
\label{eq:OmegaPExp}
\im S_0 &= \ 2\pi \, \frac{\rH}{d} \int_{0}^{\omega} \dd{\omega'} \left( 1-\frac{\omega'}{d \, M} -\frac{\textcolor{blue}{(d-1)}\, \left(\omega'\right)^2}{2 \, d^2 \, M^2} - \cdots\right) \ , \\
\label{eq:OmegaPExp2}
&= \ 2\pi \, \frac{r_{\text{H}}}{d} \int_{0}^{\omega} \dd{\omega'} \left[ 1 + \sum_{k=1}^{\infty} \frac{1}{k} \left( \frac{\omega'}{d \, M}\right)^k\times \prod_{j=0}^{k-1} (j \, d-1)  \right] \ 
\end{align}
At this moment, we pause to address the coefficient of $(\omega')^{2}$ in \eqref{eq:OmegaPExp}, which vanishes at $d = 1$. Indeed, from \eqref{eq:OmegaPExp2} we see that all terms $(\omega')^{k \geq 2}$ vanish at $d = 1$, or equivalently in $D = 4$. While this may be understood as a straightforward consequence of the binomial expansion, we also see that this truncation is special to four-dimensional physics; when $d = 1$, \emph{all} higher-order corrections vanish identically. This may be traced back to the fact that in $D = 4$ the mass of a black hole and its horizon radius are related in a linear fashion. In general, leading order backreaction may be arranged to take the form of an infinite series in $\omega$, as evinced by \eqref{eq:OmegaPExp}. The leading order result (which neglects backreaction) may be used to read off the Hawking temperature
\begin{equation}
T_{\text{BH}} = \frac{d}{4\pi \rH} \ ,
\end{equation}
consistent with known results for higher-dimensional black holes \cite{Myers:1986un}.

\subsection{Subleading Corrections in Large D}
\label{sec:SubLargeD}
Before evaluating the integral exactly, let us appeal to the limit in which $D \rightarrow \infty$. Further, in this section we choose to only work up to subleading order in the $1/D$ expansion---that is, terms of O$(1/D^2)$ are neglected---allowing us to make the identification
\begin{equation}
\frac{1}{d} \simeq \frac{1}{D}
\end{equation}
in \eqref{eq:OmegaPExp2}. To do this, we introduce the scale $\ell = r_{\text{H}}/D$ which fixes the leading divergence, and consider the infinite dimensional limit. This \emph{is} the thickness of the membrane. Since we are working up to subleading order, we keep $1/D$ corrections to the leading result, which itself is O$(\ell)$. It is satisfying to see that the subleading terms (in the $1/D$-expansion) simplify considerably:
\begin{align}
\im S_0 &\simeq \ 2\pi \, \left(\frac{r_{\text{H}}}{D} \right) \, \int_{0}^{\omega} \dd{\omega'} \left( 1-\frac{\omega'}{D \, M} -\frac{(\omega')^2}{2D \, M^2} - \cdots\right) \ , \\
&\simeq \ 2\pi \ell \int_{0}^{\omega} \dd{\omega'} \left( 1-\frac{1}{D} \sum_{k = 1}^{\infty} \frac{1}{k} \left(\frac{\omega'}{M}\right)^k \right) \ , \\
&\simeq \ 2\pi \ell \int_{0}^{\omega} \dd{\omega'} \left( 1+\frac{1}{D} \log \left( 1-\frac{\omega'}{M}\right) \right) \ .
\end{align}

For just a moment, let us neglect subleading terms, which would correspond to dropping the logarithmic piece in the above equation. This gives us
\begin{align}
\im S_0 &\simeq \ 2\pi\ell \, \int_{0}^{\omega} \dd{\omega'}  \ , \\
&\simeq \ 2\pi\ell \, \omega  \ .
\end{align}
Correspondingly, the tunneling rate evaluates to
\begin{equation}
\Gamma \sim \exp \left( -4\pi\ell \, \omega \right) \ ,
\end{equation}
allowing us to read off a temperature that is set by the thickness $\ell$:
\begin{equation}
\label{eq:LargeDTemp}
T_{\text{BH}} = \frac{1}{4\pi \ell} \ .
\end{equation}
At this point, the attentive reader might ask: in the large-$D$ limit, isn't \eqref{eq:LargeDTemp} just plain divergent? It is, and this is consistent with the fact that radial gradients of the gravitational potential diverge near the horizon \citep{Emparan:2013moa}. We find it conceptually clearer to express all thermodynamic quantities associated with a large-$D$ black hole in terms of its thickness.

Of particular interest is the fact that when we chose to neglect subleading (in $1/D$) terms, we were left with no corrections due to backreaction like the ones we saw in the case of $D = 4$. We see here that neglecting backreaction appears to go hand in hand with neglecting the subleading (in $1/D$) corrections to the imaginary part of the action, thus providing us with a concrete realization of our guiding intuition that in the limit $D \rightarrow \infty$, backreaction is suppressed in the inverse-dimensional expansion. 

Proceeding with this line of thought: what if we didn't want to neglect backreaction, i.e.~keep the subleading term too? In this case too, we find on introducing the notation $m = M - \omega$ that:
\begin{equation}
\label{eq:ImS1D}
\im S_0 \simeq 2\pi\ell \left[ \omega - \frac{1}{D} \left( \omega + m \log \frac{m}{M} \right) \right] \ .
\end{equation}
This in turn means the tunneling rate is
\begin{equation}
\label{eq:GammaSub}
\Gamma \sim \exp \left\lbrace - 4\pi\ell \, \omega + \frac{4\pi\ell}{D} \left( \omega + m \log \frac{m}{M}\right)\right\rbrace \ ,
\end{equation}
We emphasize that this result is \emph{exact} in the energy of the Hawking quanta, but only correct up to subleading order in the large-$D$ expansion. Also, the effective temperature has an expression that is cumbersome but it is easy to check that (consistent with expectations) the temperature of the black hole rises after radiation. That is, as the large-$D$ black hole radiates, it becomes hotter.

Finally, what if the particle carried away \emph{all} the mass of the black hole? Once again, we invoke the arguments of Parikh-Wilczek, who argue that there can be only one such state. By the principles of statistical mechanics, a black hole with entropy $S_{\text{BH}}$ would be expected to have a total of $e^{S_{\text{BH}}}$ states. Putting these together, the probability for finding a state which has the same mass as the black hole is $e^{-S_{\text{BH}}}$. What this means for us is that we have to set $\omega = M$, so $m \rightarrow 0$ in \eqref{eq:ImS1D}. This limit is smooth, and we get
\begin{equation}
\im S_0 \simeq 2\pi\ell \left[ M - \frac{M}{D} \right] \ ,
\end{equation}
which in turn will determine
\begin{equation}
\Gamma \sim \exp \left\lbrace -4\pi\ell \left( M - \frac{M}{D} \right) \right\rbrace = \exp (-S_{\text{BH}}) \ ,
\end{equation}
or
\begin{equation}
S_{\text{BH}} = 4\pi\ell \, \left( M - \frac{M}{D} \right) \ .
\end{equation}
The above result includes a subleading correction to the Bekenstein-Hawking entropy in the large-$D$ limit. This form is consistent with the findings of \cite{Emparan:2013moa}, who via scaling arguments find that the entropy of black holes in the strict large-$D$ limit is linear in its mass. This is quite striking: it means that there is no entropic gain when a black hole cleaves or merges with another black hole, indicating that this process is reversible in an infinite number of dimensions. This is a reflection of the extremely weak, diluted nature of the gravitational interaction in the large-$D$ limit.

\section{All Subleading Corrections}
\label{sec:kSubLead}

It is easy to see that the infinite sum \eqref{eq:OmegaPExp2} can be brought into closed form. A discussion of how this is done if one insists on working order-by-order in an inverse-dimensional expansion in the large-$D$ limit is relegated to Appendix \ref{sec:AppA}. The final answer for the imaginary part of the action for all finite-dimensional black holes is just \eqref{eq:rHMomega} with a factor of $\rH$ pulled out: 
\begin{equation}
\label{eq:SummedForm}
\im S_0 = 2\pi \, \frac{\rH}{d} \int_{0}^{\omega} \dd{\omega'} \left( 1-\frac{\omega'}{M}\right)^{1/d} \ ,
\end{equation}
and on integrating, we get
\begin{equation}
\im S_0 = 2\pi \, \rH \left[ M - m \left( \frac{m}{M} \right)^{1/d} \right] \times \frac{1}{1+d} \ ,
\end{equation}
where the above expression should be understood as fully accounting for leading-order backreaction. We can, as before, ask what the entropy of a higher dimensional black hole is. Just like earlier sections, all we need to do is let $m \rightarrow 0$. Once again, the limit is well-defined, and we get
\begin{align}
\label{eq:FiniteDEntropy}
S_{\text{BH}} &= 4\pi \rH \left( \frac{1}{1+d} \right) M \ ,
\end{align}
We can quickly confirm that this result is correct by showing that it is equivalent to the classic relation between entropy and area by using the linear relation between mass and area in arbitrary dimensions, straightforwardly derived from \eqref{eq:2SphereArea}:
\begin{equation}
\frac{16 \pi}{d+1} \, M = \frac{A}{\rH} \ .
\end{equation}
This gives us
\begin{align}
S_{\text{BH}} = \frac{A}{4} \ .
\end{align}

\section{The Large-D Limit}
\label{sec:LargeD}
The results of previous sections are applicable in any finite number of dimensions. In this section, we address the limit $D \rightarrow \infty$.

We pause here to make a point regarding the continuity of our analysis in spacetime dimension: we continue to define thickness in the same way for finite-dimensional black holes as well. Of course, it is important to note that any localization of curvature remains solely a property of the large-$D$ limit. Rather, for arbitrary dimension, finite or infinite, we adopt the point of view that this thickness is just the natural length scale in the problem:
\begin{itemize}
    \item For finite dimensional black holes, the thickness is of the same order as the horizon radius, while
    \item for infinite-dimensional black holes, the thickness goes to zero, effectively localizing any non-zero curvature effects to within O$(\ell)$ of the horizon.
\end{itemize}
We emphasise that we are \emph{not} appealing to any separation of scales in a finite number of dimensions. Indeed, in $D = 4$, the horizon radius and thickness coincide. 

\subsection{Entropy for Large-D Black Holes}
For an infinite dimensional black hole, naively taking the limit $d \rightarrow \infty$ will give us a divergent answer. We have learned from experience that as with the temperature, we should express the entropy in terms of the length scale $\ell$. It is straightforward to determine that
\begin{align}
S_{\text{BH}} &= 4\pi\ell \left(\frac{d}{1+d} \right) M \ ,
\end{align}
which for the same reasons as the finite-dimensional case is consistent with the entropy-area relation. When written as as inverse-dimensional expansion, we find:
\begin{equation}
S_{\text{BH}} = 4\pi \ell \, M \left( 1 - \frac{1}{d} + \frac{1}{d^2} + \cdots \right) \ ,
\end{equation}
That is, the black hole entropy is given as an expansion in $1/d$ off its leading order value of $4\pi\ell\,M$.\footnote{It is amusing to note that when one uses this formula for $d = 1$, the resulting is Grandi's well-known divergent series
\begin{equation*}
1-1+1-1+ \cdots \ .
\end{equation*}
While this may seem problematic, the series in question can be Borel resummed, and this procedure yields the answer $1/2$. This gives the correct answer for the black hole entropy in $d = 1$ or equivalently, $D = 4$.} This is interesting because if one neglected backreaction entirely and computed the entropy, this is precisely the answer one would get.

We take this to mean that in the leading large-$D$ limit, backreaction can be safely neglected, and $1/d$ corrections incorporate the effects of backreaction. This is consistent with our expectation that backreaction and the inverse-dimensional expansion are naturally paired. Finally, as we have already mentioned, it is easy to verify that the entropy density is \emph{always} maintained at the value $1/4$. At leading order, this is consistent with the findings of \cite{Bhattacharyya:2016nhn}.

\section{Conclusions}
\label{sec:Conclusions}
We have computed the Bekenstein-Hawking entropy for Schwarzschild-Tangherlini black holes in all dimensions---finite and infinite---and confirmed the validity of our expressions by checking that the entropy-area relation is satisfied. An up-shot of this analysis is that we are able to treat both finite- and infinite-dimensional black holes in the same picture, appealing in each case to length scales native to the system. We expect that, following \cite{Parikh:1999mf}, similar results should hold for the Reissner-Nordstr\"om black hole. We now turn to some directions for future research.

First, in Appendix \ref{sec:AppA}, we learned that the ``right'' way to take the large-$D$ limit was to send $d \rightarrow \infty$ and not $D \rightarrow \infty$ as we might have naively thought. In particular, we saw that the coefficients in the $1/d$ expansion were easier to identify and resum than those in the $1/D$ expansion. This is due to mixing beyond subleading order in the $1/D$ expansion. The number $d$ in our analysis parametrizes an infinite family of static, spherically symmetric solutions to the Einstein field equations, so it would be interesting to see whether this is true for other families of solutions as well. 

Second, an obvious extension of these techniques would be to determine subleading (in $\hbar$) corrections, which are known to be logarithmic in the area. An all-orders WKB ansatz does in fact reaffirm this. However, within the tunneling framework such logarithmic corrections in the exponential would be sensitive to constants multiplying the tunneling probability. As it stands, we read off the entropy of the black hole by looking at what is sitting in the exponential of the tunneling probability, paying no attention to any constants that may multiply it. This will require a more careful analysis that accounts for both positive and negative frequency modes.

Finally, it is often claimed that the large-$D$ limit of general relativity is similar to the large-$N$ limit in Yang-Mills theories. That is, just as we hope to understand Yang-Mills at $N = 3$ by studying Yang-Mills in the limit of infinitely many colors, we also hope to better understand gravity in $D = 4$ by studying gravity in the limit of infinitely many dimensions. In order to explore the extent to which this analogy can be pushed, it would be interesting to explore the double scaling limit
\begin{equation}
D \rightarrow \infty \quad \text{and} \quad \hbar \rightarrow 0 \ ,
\end{equation}
while keeping some combination of the spacetime dimension and Planck's constant fixed. Once again, this will require us to compute quantum actions for tunneling. We hope to return to this task in the near future.

\appendix

\section{Each Order in Large-D}
\label{sec:AppA}
Bolstered by the success of the subleading computation in Section \ref{sec:SubLargeD}, we now see if we get the same answer if we insist on working order-by-order in the $1/D$ expansion.
\subsection{2- and 3-Subleading Corrections}
At $2$- and $3$-subleading orders, for example, we have the contribution (using the shorthand $x = \omega/M$)
\begin{equation}
\label{eq:D2x}
-\frac{1}{D^2} \left( 3\, x + x^2 + \frac{1}{2} \,x^3 + \frac{7}{24} \,x^4 + \frac{11}{60}\, x^5 + \frac{43}{360}\, x^6 + \cdots \right) \ ,
\end{equation}
\begin{equation}
\label{eq:D3x}
-\frac{1}{D^3} \left( 9\, x + \frac{3}{2} \,x^2 + \frac{1}{6} \,x^3 - \frac{1}{4} \,x^4 - \frac{49}{120} \,x^5 - \frac{113}{240} \,x^6 - \cdots \right) \ .
\end{equation}
In short, $k$-subleading expansions are straightforwardly computed, but their resummation is unclear at the moment. Our investigations have led us to conclude that this is because of collecting the above expansions in the ``wrong'' dimension of spacetime. We correct this error in the following section, and in the process are able to determine how this resummation is done for all spacetime dimensions, finite or infinite.

\subsection{All Subleading Corrections}
\label{sec:AllSubLead}
The goal of this section is to show that each $k$-subleading correction to the entropy of a black hole in any dimension $D \geq 4$ has a series that can be resummed. This resummation will reveal some interesting relations to the study of permutations and cycles.

We go back to the expression \eqref{eq:OmegaPExp2}, which is valid in all spacetime dimensions, and choose to make the dependence on $D$ explicit:
\begin{align}
\im S_0 &= \ 2\pi \, \frac{r_{\text{H}}[M]}{D-3} \int_{0}^{\omega} \dd{\omega'} \left[ 1 + \sum_{k=1}^{\infty} \frac{x^k}{k \, (D-3)^k} \times \prod_{j=0}^{k-1} (j(D-3)-1)  \right] \ .
\end{align}
Now, let us focus on this expression at each order in $x = \omega'/M$. We list the coefficients of $x^k$ in the above expression for small values of $k$:
\begin{align}
\nonumber k=1 : \quad -&\frac{1}{(D-3)} \ , \\
\nonumber k=2 : \quad -&\frac{(D-4)}{2(D-3)^2} \ , \\
k=3 : \quad -&\frac{(D-4)(2D-7)}{3(D-3)^3} \ , \\
\nonumber k=4 : \quad -&\frac{(D-4)(2D-7)(3D-10)}{4(D-3)^4} \ , \\
\nonumber k=5 : \quad -&\frac{(D-4)(2D-7)(3D-10)(4D-13)}{5(D-3)^5} \ ,
\end{align}
and so on. This is not the best way to write the above expressions, however, as there is an underlying structure that is not manifest in this presentation. So let us instead decompose each of these expressions into partial fractions. This gives us:
\begin{align}
\nonumber k=1 : \quad -&\frac{1}{d} \ , \\
\nonumber k=2 : \quad +&\frac{1}{2d^2} - \frac{1}{2d} \ , \\
k=3 : \quad -&\frac{1}{6d^3} + \frac{1}{2d^2} - \frac{1}{3d}\ , \\
\nonumber k=4 : \quad +&\frac{1}{24d^4} - \frac{1}{4d^3} + \frac{11}{24d^2} - \frac{1}{4d} \ , \\
\nonumber k=5 : \quad -&\frac{1}{120d^5} + \frac{1}{12d^4} - \frac{7}{24d^3} + \frac{5}{12d^2} - \frac{1}{5d} \ ,
\end{align}
which is a little better. This presentation makes it easier to see that terms contributing to a specific order $k$ in leading backreaction will only contribute up to $O(d^{-k})$ in the inverse-dimensional expansion. It also suggests that perhaps expanding in $d=D-3$ instead of $D$ makes more sense, which is consistent with the findings of \cite{EmparanTalk} up to a numerical factor. Let's do one final thing. We observe that the coefficient of the \emph{largest} inverse power of $d$ at each level is an inverse factorial. If we insist that the leading coefficient be unity---this is often a desirable thing---we must multiply each coefficient of $x^k$ by $k!$, keeping in mind that we'll need to compensate for this factor in the denominator soon. Effecting this multiplication, we get
\begin{align}
\nonumber k=1 : \quad -&\frac{1}{d} \ , \\
\nonumber k=2 : \quad +&\frac{1}{d^2} - \frac{1}{d} \ , \\
\label{eq:Array3}
k=3 : \quad -&\frac{1}{d^3} + \frac{3}{d^2} - \frac{2}{d}\ , \\
\nonumber k=4 : \quad +&\frac{1}{d^4} - \frac{6}{d^3} + \frac{11}{d^2} - \frac{6}{d} \ , \\
\nonumber k=5 : \quad -&\frac{1}{d^5} + \frac{10}{d^4} - \frac{35}{d^3} + \frac{50}{d^2} - \frac{24}{d} \ ,
\end{align}
and so on. We make two observations regarding the structure of the above coefficients. First, the coefficients of $1/d$ are $(k-1)!$, which hints to us that the above array of coefficients should be read diagonally---collecting the series in inverse powers of $d$, that is---and also that perhaps some generalization of the ``number of permutations'' explains the structure of the other diagonals.

Pick $k = 3$, and read the absolute value of the coefficients---this corresponds to reading the array of coefficients in \eqref{eq:Array3} horizontally. Of the total number of permutations of three elements (in this case $6$):
\begin{itemize}
    \item the number of permutations with $3$ cycles is $1$, the identity permutation: $(1)(2)(3)$,
    \item the number of permutations with $2$ cycles is $3$: $(1)(23)$, $(3)(12)$, and $(2)(13)$,
    \item the number of permutations with $1$ cycle is $2$: $(132)$ and $(123)$.\footnote{We are using cycle notation here. For example, the cycle $(132)$ means we map $1 \rightarrow 3$, $3 \rightarrow 2$, and $2 \rightarrow 1$. Similarly, disjoint cycles like $(1)(23)$ means $1 \rightarrow 1$, $2 \rightarrow 3$, and $3 \rightarrow 2$.}
\end{itemize}
With a little more work, one can show that at level $k$, the total number of permutations of all elements is $k!$, and of those permutations, the number of permutations with $j$ cycles is the absolute value of the coefficient of $1/d^j$.

Let us write out the array of coefficients in a more suggestive manner.
\begin{equation}
\label{eq:StirlingArray}
\begin{matrix}
\violet{-1} \\ \\
\cyan{+1} & \violet{-1} \\ \\ 
\green{-1} & \cyan{+3} & \violet{-2} \\ \\
\blue{+1} & \green{-6} & \cyan{+11} & \violet{-6} \\ \\
\orange{-1} & \blue{+10} & \green{-35} & \cyan{+50} & \violet{-24} \\ \\
\red{+1} & \orange{-15} & \blue{+85} & \green{-225} & \cyan{+274} & \violet{-120} \\ \\
-1 & \red{+21} & \orange{-175} & \blue{+735} & \green{-1624} & \cyan{+1764} & \violet{-720} \\ \\
\olive{+1} & -28 & \red{+322} & \orange{-1960} & \blue{+6769} & \green{-13132} & \cyan{+13068} & \violet{-5040}  
\end{matrix}
\end{equation}
In the above equation, one should imagine two axes:
\begin{itemize}
    \item the $j$-axis runs horizontally, each giving the coefficient of a different inverse power of $d$ at a fixed order $(x^k)$ in the $x$-expansion, and
    \item the $k$-axis runs along the colour lines---from north-west to south-east---and each contributing to the \emph{same} order in the $1/d$-expansion, but across orders in the $x$-expansion!
\end{itemize}
These numbers---the number of permutations of $k$ elements that have $j$ cycles---are well studied in the literature on combinatorics \cite{Graham:1994}. The resultant sequences of numbers coloured the same (and neglecting their signs!) are denoted $S(k,j)$, the \emph{unsigned Stirling numbers of the first kind}. Since we have been looking for a natural inverse-dimensional expansion in which to collect terms of the same order in $1/d$, it is natural that we read this array diagonally, along the $k$-axis.

Further (on remembering that we still have a $k!$ in the denominator) it is easy to show that 
\begin{equation}
\sum_{k = 0}^{\infty} \frac{S(k,j)}{k!} \, x^k = \frac{\log^j \,(1-x)}{j!} \ .
\end{equation}
We can use this identity to resum---at each $k$-subleading order in the $1/d$-expansion---the series in $x = \omega/M$. As a preliminary check, the case $j = 1$ which concerns itself with the outermost diagonal in \eqref{eq:StirlingArray} reproduces the subleading results we derived earlier when we were expanding in $1/D$ instead of $1/d$. This is as it should be, because the inverse-dimensional expansions in $D$ and $d$ only differ at $2$-subleading order.

Putting all this together, we have:
\begin{equation}
\label{eq:LogForm}
\im S_0 = 2\pi \ell \int_{0}^{\omega} \dd{\omega'} \left[ \sum_{k=0}^{\infty} \frac{1}{k! \, d^k} \log^{k} \left( 1-\frac{\omega'}{M}\right) \right]\ ,
\end{equation}
where $d = D-3$. Clearly, sending $d \rightarrow \infty$ instead of $D \rightarrow \infty$ seems to make a small but notable difference in so far as the coefficients are concerned. In one expansion, the coefficients are inexplicable, while in the other expansion the underlying combinatorial structure unravels rather naturally. Armed with this perspective we now understand why the coefficients in \eqref{eq:D2x} and \eqref{eq:D3x} were so bewildering. If one takes the results in the $(1/d)$ expansion and insists on expanding them in $1/D$, then we effect the substitution
\begin{equation}
\frac{1}{d} = \frac{1}{D-3} = \frac{1}{D} \sum_{k = 0}^{\infty} \frac{3^k}{D^k} \ ,
\end{equation}
and we see that the coefficient of (say) $1/D^2$ receives contributions from all the diagonals in \eqref{eq:StirlingArray}! It appears to be the case that choosing to expand in $1/d$ \emph{crystallizes} the inverse-dimensional expansion, while expanding in $1/D$ seems to make the inverse-dimensional expansion more amorphous.\footnote{We mean this quite literally: the $1/d$ expansion allows for an array of coefficients that do not mix with each other. This lattice-like structure is destroyed if one chooses to expand in powers of $1/D$.} Thus, we have shown that the results quoted in Section \ref{sec:kSubLead} are true even in the limit $D \rightarrow \infty$. Finally, performing the sum in \eqref{eq:LogForm} yields \eqref{eq:SummedForm}.

\bibliographystyle{JHEP}
\bibliography{Refs}
\end{document}